\documentclass[conference]{IEEEtran}
\IEEEoverridecommandlockouts
\usepackage{cite}
\usepackage{amsmath,amssymb,amsfonts}
\usepackage{physics}
\usepackage{algorithmic}
\usepackage{graphicx}
\usepackage{textcomp}
\usepackage{xcolor}
\usepackage{multirow}
\def\BibTeX{{\rm B\kern-.05em{\sc i\kern-.025em b}\kern-.08em
    T\kern-.1667em\lower.7ex\hbox{E}\kern-.125emX}}
\begin{document}

\title{Extremum seeking control of quantum gates}

\author{\IEEEauthorblockN{Erfan Abbasgholinejad}
\IEEEauthorblockA{\textit{University of Washington, Seattle}}
\and
\IEEEauthorblockN{Haoqin Deng}
\IEEEauthorblockA{\textit{University of Washington, Seattle}}
\and
\IEEEauthorblockN{John Gamble}
\IEEEauthorblockA{\textit{IonQ, Inc.}}
\and
\IEEEauthorblockN{J. Nathan Kutz}
\IEEEauthorblockA{\textit{University of Washington, Seattle}}
\and
\IEEEauthorblockN{Erik Nielsen}
\IEEEauthorblockA{\textit{IonQ, Inc.}}
\and
\IEEEauthorblockN{Neal Pisenti}
\IEEEauthorblockA{\textit{IonQ, Inc.}}
\and
\IEEEauthorblockN{Ningzhi Xie}
\IEEEauthorblockA{\textit{University of Washington, Seattle}}}

\maketitle

\begin{abstract}
To be useful for quantum computation, gate operations must be maintained at high fidelities over long periods of time.
In addition to decoherence, slow drifts in control hardware leads to inaccurate gates, causing the quality of operation of as-built quantum computers to vary over time.
Here, we demonstrate a \emph{data-driven} approach to stabilized control, combining extremum-seeking control (ESC) with direct randomized benchmarking (DRB) to  stabilize two-qubit gates under unknown control parameter fluctuations.
As a case study, we consider these control strategies in the context of a trapped ion quantum computer using physically-realistic simulation. 
We then experimentally demonstrate this control strategy on a state-of-the-art, commercial trapped-ion quantum computer.
\end{abstract}

\section{Introduction}
In order to realize the promise of quantum computing, researchers are investigating a wide variety of hardware technology platforms.
In all of these systems, the as-built processor suffers imperfections that impact performance. 
Some of these problems occur due to quantum effects intrinsic to the qubits, such as decoherence.
However, in many platforms, imperfections in the \emph{classical} control dominate the error budget, entering as either coherent errors (\emph{e.g.}, over-rotating due to an imperfectly calibrated control gain) or as incoherent errors (\emph{e.g.}, due to stochastic noise coupling through the control fields to the qubit).
Eliminating these technical errors is thus imperative for reliable quantum computation.

Increasing the performance of quantum operations via control stabilization has a long history, spanning the multitude of physical architectures currently being pursued. 
Open-loop strategies have been employed since Nuclear Magnetic Resonance systems to design robust gate pulses \cite{khaneja2005optimal}, and these techniques have been adapted to superconductors \cite{motzoi2009simple}, trapped ions \cite{blumel2021power}, and spin qubits \cite{frees2019adiabatic}, among others. 
In addition, closed-loop control strategies have been investigated for tuning QPU performance \cite{egger2014adaptive,kelly2014optimal,ferrie2015robust}.

As qubit performance improves, this optimization becomes more challenging -- the models required for most closed- and open-loop strategies need to be high enough fidelity to resolve small deviations from ideal behavior.
Models, especially those which capture the impact of experimental control parameters on qubit gate performance, are not perfect.
Hence, model-based approaches can become dominated by out-of-model errors, or else require prohibitively large model training times.
The alternative is to use a model-free approach, but this also comes with downsides. 
In particular, model-free approaches need to construct gradients of the local objective function landscape empirically and with high precision.
This procedure is intensive in the number of required experiments, and often does not scale well in the number of control parameters.

Here, we employ Extremum Seeking Control (ESC) \cite{ariyur2003real}, a widely-used model-free control strategy that is both simple and has favorable scaling with control space dimensionality.
We combine ESC with direct randomized benchmarking (DRB) \cite{proctor2019direct} and demonstrate a model-free closed-loop control paradigm. 
As a case-study, we consider a trapped ion system with three physical control parameters, demonstrating that ESC works in both simulation and experiment. 
In practice, the ESC strategy could be deployed alongside efficient model-based feedback strategies as an error polishing step to eliminate residual out-of-model contributions to gate infidelity, instead of as a standalone feedback controller.

\begin{figure}[tb]
\centerline{\includegraphics[width=0.5\textwidth]{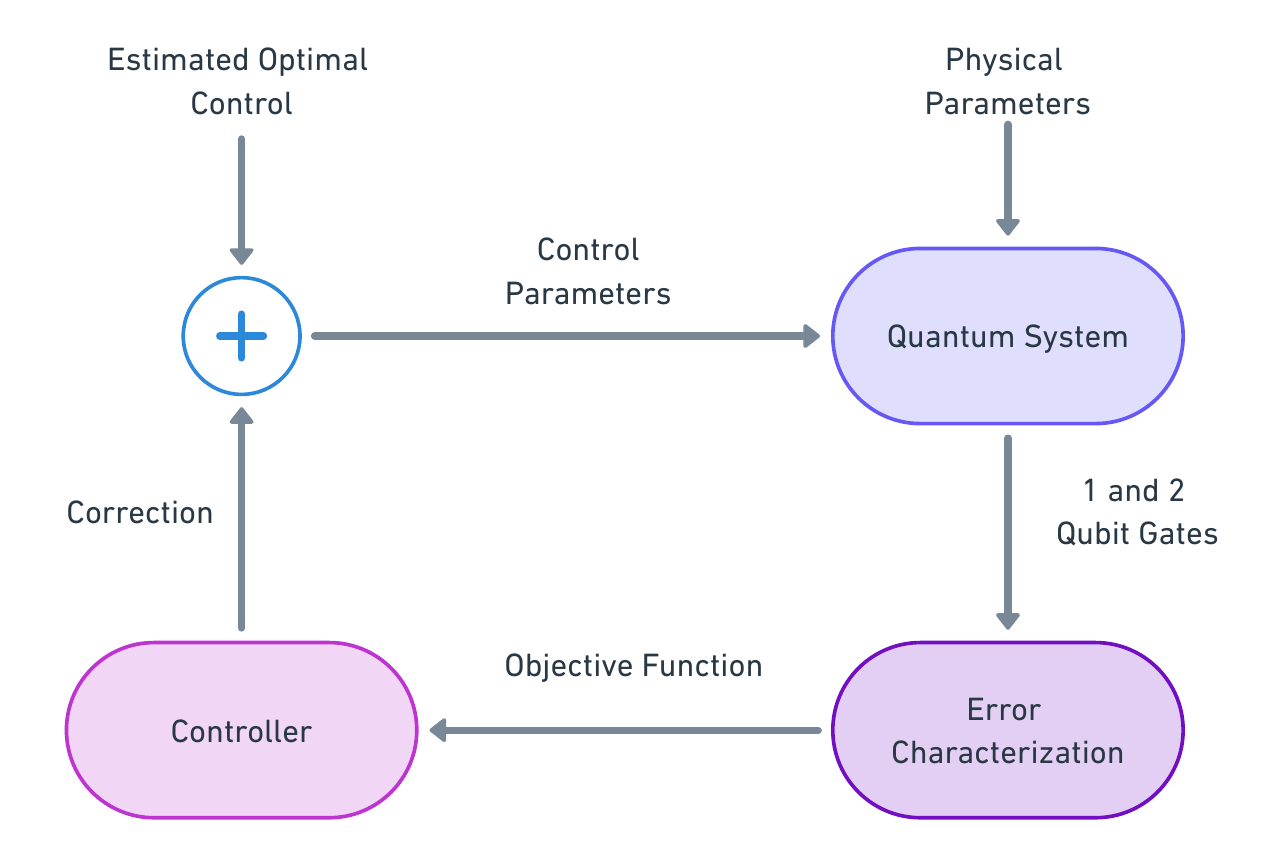}}
\caption{\label{fig:block_diagram}
Block diagram of the closed-loop control employed in this work.
The quantum system (either a simulator or a physical system), takes as input control parameters and time varying physical parameters. 
We use quantum characterization (DRB in our case) to learn an objective function and feed this to the controller.
The controller then determines a correction to the control parameters, feeding this back to the quantum system.}
\end{figure}

This work is organized as follows.
In Sec.~\ref{sec:control_strat}, we define the control problem and discuss how ESC works.
We then present a case-study in Sec.~\ref{sec:sim_results}, simulating ESC on a trapped-ion quantum computer with realistic parameter drift.
Sec.~\ref{sec:exp_results} shows a demonstration of this method in experiment, and Sec.~\ref{sec:conclusion} offers concluding remarks and future directions.

\section{Control strategy}\label{sec:control_strat}

Our general control strategy is shown in Fig.~\ref{fig:block_diagram}.
The objective function studied in this work is the gate fidelity, which is not directly observable but can be measured indirectly via standard quantum characterization methods. We use direct randomized benchmarking (DRB) for this purpose.

\begin{figure}[tb]
\centerline{\includegraphics[width=0.5\textwidth]{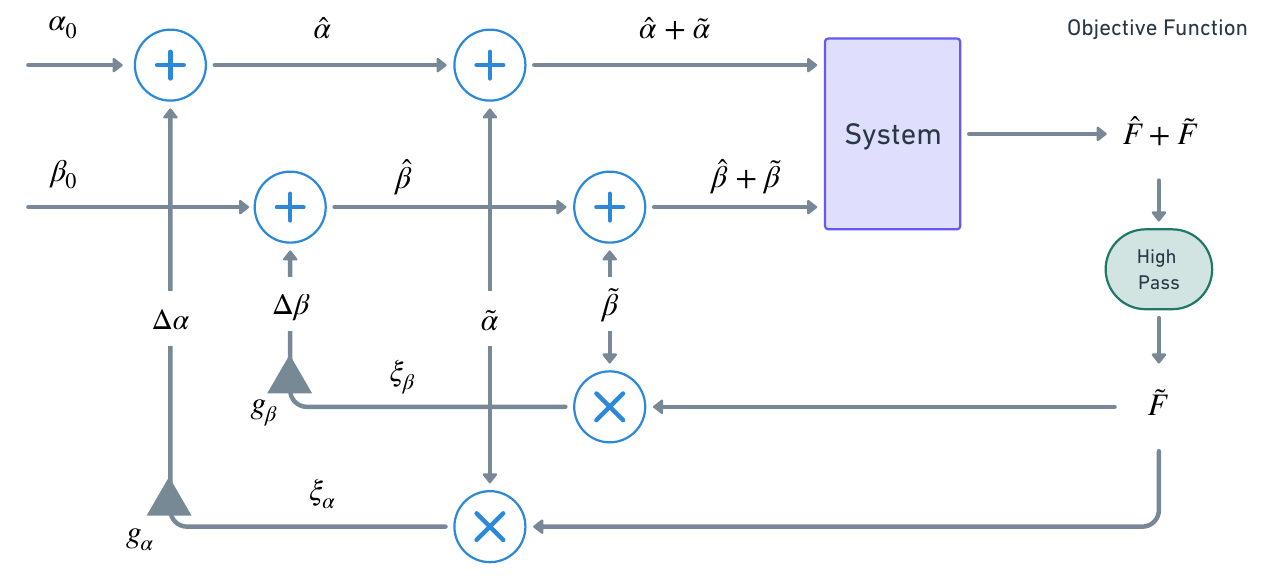}}
\caption{\label{fig:esc}
Control diagram for ESC.
As explained in the main text, ESC functions by frequency multiplexing deliberate control perturbations to approximate the local derivative of an objective function.
Unlike more complex optimization techniques, ESC is extremely simple to implement and analyze (either in software or hardware).}

\end{figure}

We use an ESC controller \cite{ariyur2003real} to stabilize the control parameters against physical drifts, shown schematically in Fig.~\ref{fig:esc}, by optimizing the DRB objective function $\hat F$.
ESC starts with a set of control parameters which are assumed to be near the optimal point, noted as $\alpha_0,\,\beta_0$ in Fig.~\ref{fig:esc}. Small sinusoidal perturbations $\Tilde{\alpha},\,\Tilde{\beta}$ with distinct frequency and phase are added to the starting control parameters:
\begin{align}
\Tilde{\alpha}_i &= A_{\alpha}\sin{(\omega_{\alpha}t_i + \phi_{\alpha})}, \\
\Tilde{\beta}_i &= A_{\beta}\sin{(\omega_{\beta}t_i + \phi_{\beta})},
\end{align}
where $t_i$ ranges uniformly from 0 to 1, and $N_t$ denotes the number of points.
The perturbations in control parameters will result in a deliberate disturbance of the objective function, $\Tilde{F}$, which is separated from slow drift via a high pass filter.
This filtered signal is demodulated by integrating against each control parameter perturbation signal to obtain $\xi$, which indicates the local derivative of the corresponding control parameter. 
These local derivatives are multiplied by gain factors $g$ to obtain corrections for the control parameters, $\Delta\alpha,\,\Delta\beta$:
\begin{equation}
\begin{array}{cc}
\xi_{\alpha} = \sum_{i}^{N_t} \Tilde{\alpha}_i \cdot \Tilde{F}_i / N_t, & \Delta \alpha = \xi_{\alpha} g_{\alpha}\\
\xi_{\beta} = \sum_{i}^{N_t} \Tilde{\beta}_i \cdot \Tilde{F}_i / N_t, &
\Delta \beta = \xi_{\beta} g_{\beta}\\
\end{array}.
\end{equation}
This process proceeds iteratively so that the control parameters converge to an optimal point where the objective function is maximized. 
The hyperparameters for ESC are given by $N_t$ and $\{A,\,g,\,\omega,\,\phi\}$ for each control parameter, and can be tuned to increase convergence speed and maintain robustness against noise in the objective function $\hat F$.

\section{Simulation results}\label{sec:sim_results}

\subsection{Physical model}

\begin{figure}[tb]
\centerline{\includegraphics[width=0.5\textwidth]{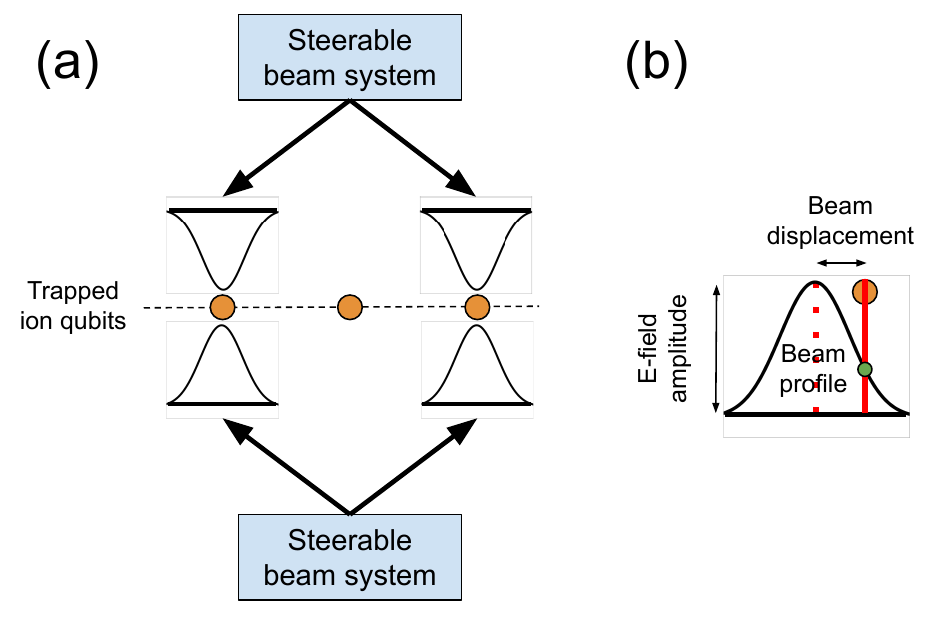}}
\caption{Schematic illustration of the experimental apparatus. Pairs of steerable beams can address each qubit, where the displacement, electric field amplitude, phase, and frequency of each beam can be controlled by synthesized RF signals sent to acousto-optic devices in the apparatus. \textbf{(inset)} The electric field amplitude of each beam evaluated at the ion position is directly proportional to the Rabi rate $\Omega$, while the phase and frequency (not shown) correspond to other parameters appearing in the atomic physics Hamiltonian. The calibration controller must stabilize each of the Hamiltonian parameters by adjusting the RF control signals under unknown, noisy drift of the physical hardware.\label{fig:control_diagram}}
\end{figure}

As a case study, we consider a simple model of a trapped-ion quantum processor.
Here, the qubit is defined by the two internal spin hyperfine levels of the atom. 
The energy eigenstates of these two levels form the computational basis of the qubit, $\{\ket{0}, \ket{1}\}$, and a two-photon Raman transition is used to coherently manipulate the qubit. 
When the frequency difference between the two Raman laser beams is near-resonant with the qubit frequency splitting, the qubit experiences Rabi dynamics~\cite{Wineland1998}. 
Individual parameters of the Hamiltonian can be controlled by adjusting the phase, frequency, and amplitude of RF control signals in the apparatus.

The native single-qubit gate is the rotation $R(\theta, \phi)$ with rotation angle $\theta$ about axis $\phi$. 
The angle $\theta = \Omega t$ is determined by the rotation time $t$ multiplied by the Rabi rate $\Omega$, which is proportional to the product of electric field amplitudes of the two Raman beams. The phase $\phi$ is determined by the relative phase difference of the two Raman beams.
The native two-qubit gate is the M\o{}lmer-S\o{}rensen (MS) gate, $MS(\chi, \phi_1, \phi_2)$~\cite{Srensen1999}. It implements a two-qubit rotation about $\sigma_{\phi_1} \otimes \sigma_{\phi_2}$ by detuning the two-photon Raman transition near the red and blue motional sidebands of the ion chain. The gate parameters $\phi_1,\,\phi_2$ are the rotation axes of the two qubits, and $\chi$ is the rotation angle.

We introduce physical hardware drift in the model via latent perturbations in the phase and amplitude of the Raman beams. To do this, we parameterize the physical state $P$ for each qubit in terms of $g2e$, the gain-to-electric-field proportionality, and $\psi^{2q}$, the phase offset that accounts for differences between the single-- and two-qubit rotation axes.  $g2e$ represents the conversion of the RF signal gain to electric field amplitude at the ion (controlling the Rabi rate), and aggregates several physical effects.

The drifting physical parameters are compensated by adjusting the RF control parameters that affect the beam alignment, amplitude, and phase, as shown schematically in Fig.~\ref{fig:control_diagram}. 
We control the RF amplitude (gain) $g$ of one of the control signals for a given beam pair, and the phase  difference between the two RF signals for a given beam pair $\psi$, which translates to an optical phase difference between the two beams.

Mathematically, we have a map from our physical and control parameters to our single-qubit gate $\{P(g2e, \psi^{2q}), C(g, \psi)\} \rightarrow R(\theta, \phi)$, defined by $\theta  \varpropto g \cdot g2e$ and $\phi = \psi$.
Similarly, for an MS gate we have two sets of physical and control parameters (one for each qubit) given by $\{P_1,C_1\}$ and $\{P_2,C_2\}$. 
The mapping $\{P_1,C_1\},\{P_2,C_2\} \rightarrow MS(\chi, \phi_1, \phi_2)$ is given by $\chi  \varpropto \left(g_1  \cdot g2e_1\right)\cdot\left( g_2  \cdot g2e_2\right)$ and $\phi_i = \psi_i + \psi_i^{2q}$.

We note that in an MS gate, many other parameters of the system can contribute to gate infidelity. 
For example, drifts in the motional mode frequencies can result in incomplete decoupling of the modes at the end of the gate. 
This particular error can be mitigated by open-loop pulse shape design~\cite{Shantanu_thesis,blumel2021power}.
The focus of this work is on the coherent gate control described above, but it can be readily combined with these other gate design approaches. 
We synthesize drifting physical parameters for a pair of qubits to simulate the real physical system based on historical data from the experiment. 
The evolution of the physical parameters over time are considered to be sinusoidal functions with randomly drifting amplitudes and frequencies $a(t) = A(t) \sin{\left(\omega(t) \cdot t\right)}$, with $\Delta A/\Delta t = \tau_a$ and $\Delta \omega / \Delta t = \tau_{\omega}$, where $\tau_a \sim \mathcal{N}(0,\sigma_a)$ and $\tau_{\omega}\sim\mathcal{N}(0,\sigma_{\omega})$

\subsection{Stabilized gate operation}

\begin{table}[tb]
\begin{center}
\begin{tabular}{ |c|c|c|c| } 
 \hline
   & \multicolumn{3}{c|}{Control parameters} \\  
  \hline
 Hyperparameters & $g_1 g_2$ & $\psi_1$ &  $\psi_2$  \\  
  \hline
 perturbation amp, A  & 0.00525 & 0.021 & 0.021  \\ 
  \hline
 perturbation freq, $\omega$ & $8\pi$ & $4\pi$ & $4\pi$  \\ 
 \hline
  perturbation phase, $\phi$ & 0 & 0 &  $\pi$ \\ 
 \hline
  gain & 10000 & 7500 & 10500 \\ 
 \hline
\end{tabular}
\newline
\caption{\label{tab:ESC_hyper_para} ESC control hyperparameters
}
\end{center}
\end{table}

\begin{table*}[tb]
\begin{center}
\begin{tabular}{ |c|c|c|c|c|c|c|c| } 
 \hline
 param set & ESC interval (min) & \# circuits per DRB depth & \# shots per circuit & \# ESC iterations & $N_t$ & runtime (min/h) & inf. suppression \\  
  \hline
 1 & 75 & 5 & 18 & 3 & 30 & 15.3 & 12.3  \\  
  \hline
 2 & 70 & 6 & 16 & 5 & 28 & 27.6 & 14.5  \\ 
  \hline
 3 & 50 & 6 & 21 & 5 & 30 & 54.3 & 21.0  \\ 
 \hline
\end{tabular}
\newline
\caption{\label{tab:hyper_para} Hyperparameter sets and their associated time costs and gate error suppression.
Here, $N_t$ is the number of ESC sample points. 
To compute the runtime cost of the calibration, we assume the single-qubit gate time to be 90~$\mu$s, and the two-qubit gate time to be 700~$\mu$s.
The final column is the infidelity suppression ratio (higher score is better).
}
\end{center}
\end{table*}

\begin{figure}[tb]
    \centering
    \includegraphics[width=0.5\textwidth]{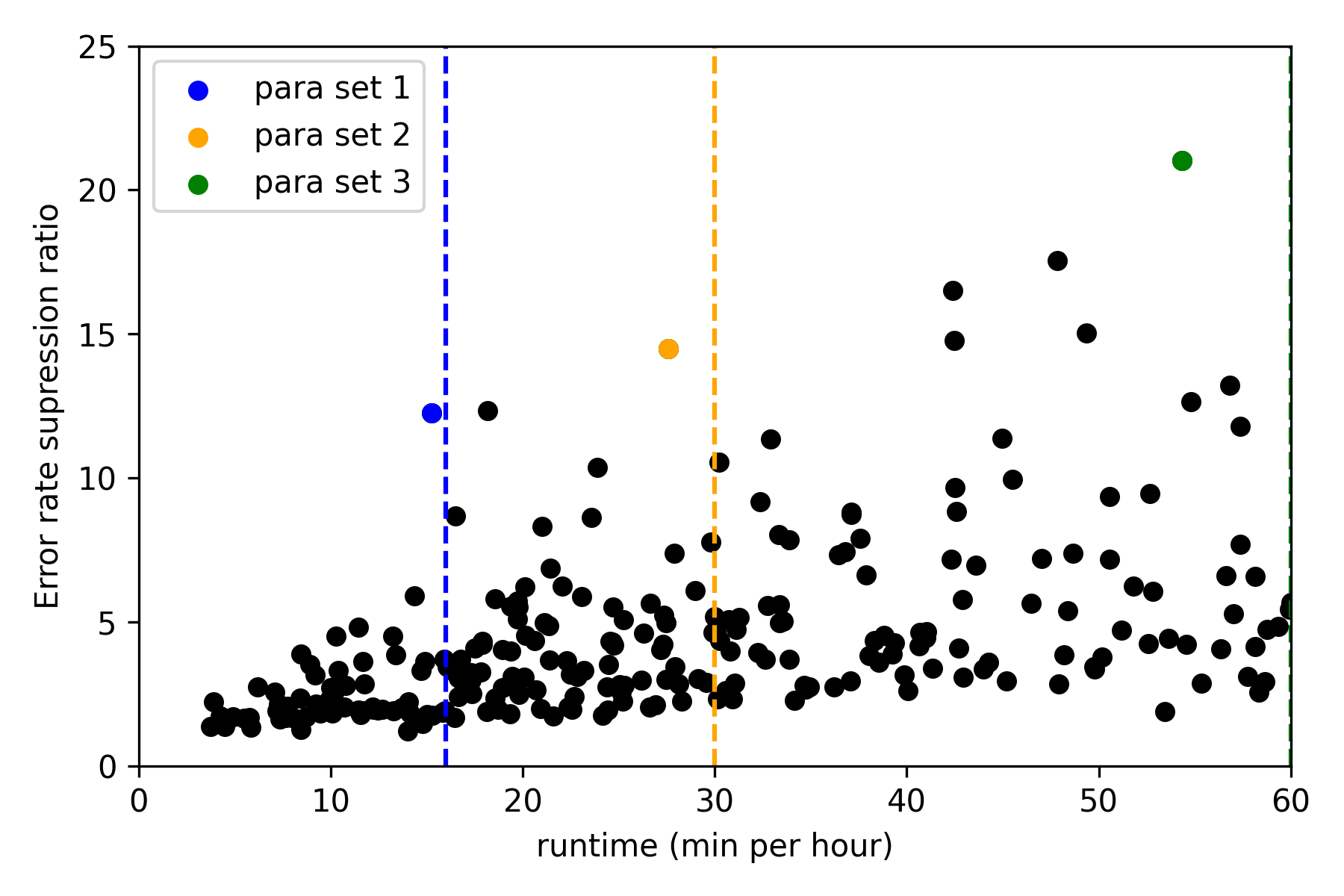}
    \caption{\label{fig:GridSearchResult} The runtime and 2Q gate error rate suppression ratio for different hyperparameter sets in the grid search.  Table~\ref{tab:hyper_para} lists the specific hyperparameter values for the three highlighted points.}
\end{figure}

\begin{figure}[tb]
    \centering
    \includegraphics[width=0.5\textwidth]{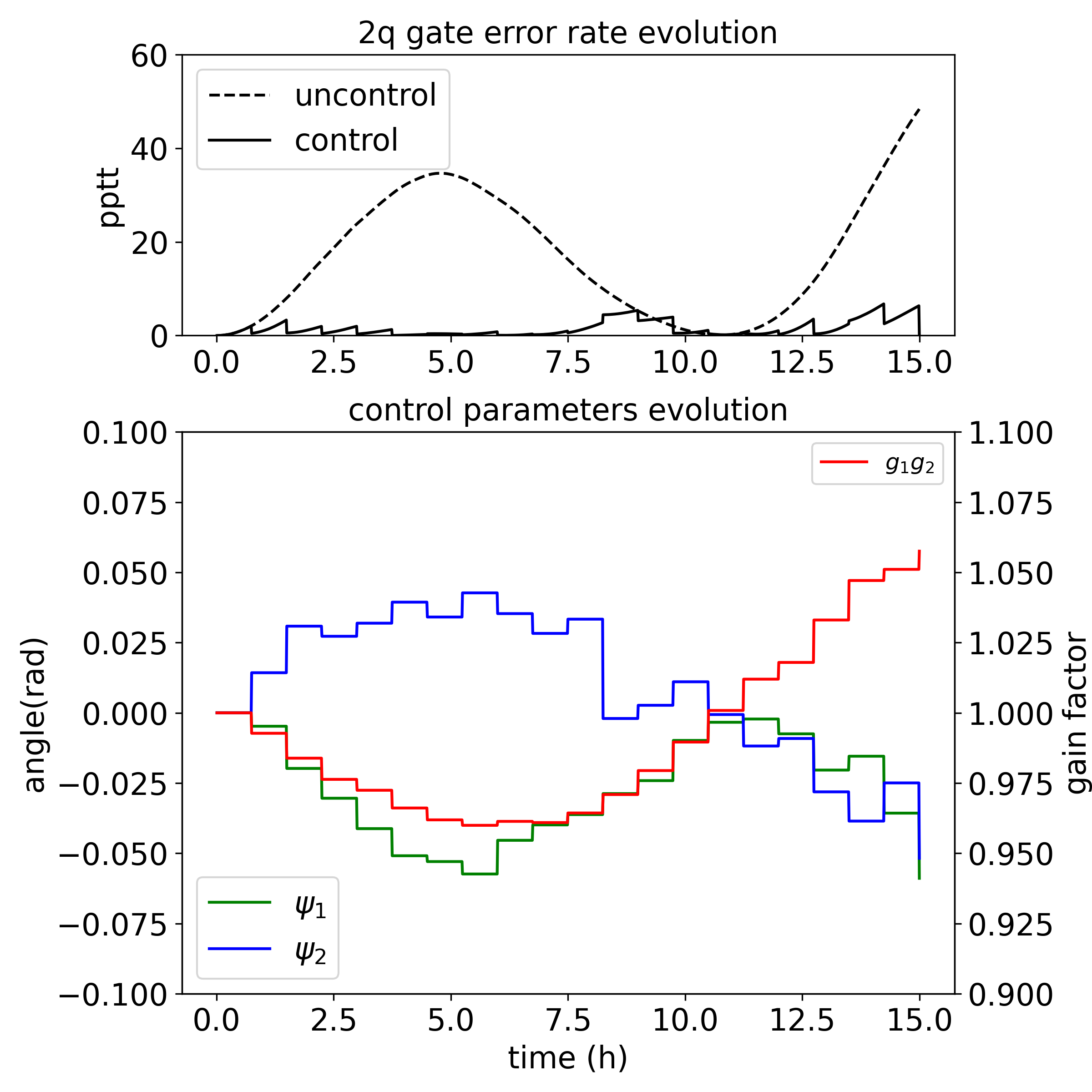}
    \caption{\label{fig:fidelity_evolution} Simulated evolution of the 2Q gate error rate and the control parameters under ESC control using hyperparameter set 1 from Table~\ref{tab:hyper_para}.}
\end{figure}

We simulate extremum-seeking control of an MS gate under the realistic noise trajectories described above.
First, we decide the hyperparameters in the ESC control, which are the amplitude, frequency, and phase of the sinusoidal perturbation and the gain factor for the variables (control parameters in the trapped ion quantum system) in the ESC control. These parameters are summarized in Table~\ref{tab:ESC_hyper_para}. Subsequently, we define the hyperparameters in the DRB, which are the number of circuits per DRB depth and the number of shots per circuit. These two hyperparameters, along with the number of sampling points in ESC perturbation signal, the number of ESC iterations per calibration, and the calibration interval, decide the runtime cost and the effectiveness of the ESC control. We perform a grid search over these hyperparameters, computing the infidelity suppression as a function of the runtime cost given a noise trajectory sampled as described in the previous section.
The results are shown in Fig.~\ref{fig:GridSearchResult}, and Table~\ref{tab:hyper_para} shows the hyperparameters for the highlighted points along with the estimated runtime. 
For these simulations, we fix an RB design consisting of depths [1, 32, 128], a two-qubit gate fraction of 0.75, and use a gate error model consisting of systematic offsets for the relevant gate parameters as determined by the noise trajectory.

We pick a candidate set of hyperparameters (set \#1) that balances the infidelity suppression with reasonable runtime cost, and in Fig.~\ref{fig:fidelity_evolution} we plot the simulated evolution of the gate fidelity under ESC over 15 hours. 
We actively control three parameters: $\psi_1$, $\psi_2$, and $g_1 g_2$, stabilizing the internal gate parameters $\phi_1, \phi_2$, and $\chi$. 
A large improvement over the uncontrolled fidelity can be seen, as the ESC controller tracks the underlying parameter drift.
In general, higher infideltiy suppression can be achieved with larger runtime overheads, and this tradeoff evaluated in the context of the experiment.

\section{Experiment}\label{sec:exp_results}
\begin{figure}
\includegraphics[width=0.5\textwidth]{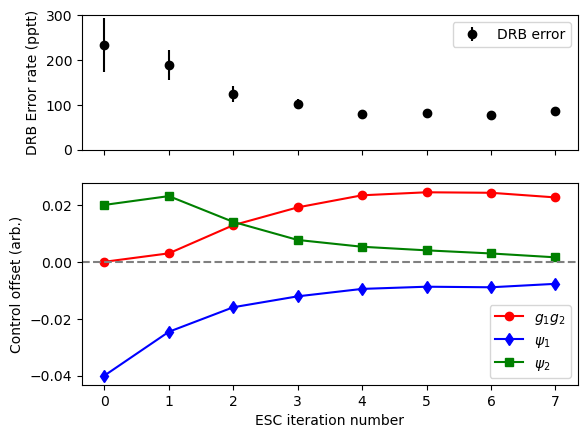}
\caption{Experimental demonstration of the ESC algorithm described in Sec.~\ref{sec:control_strat}, controlling the envelope gain $g_1 g_2 \propto \chi$, and the $\phi_1$, $\phi_2$ parameters of the MS unitary via $\psi_1,\,\psi_2$. For each iteration of the algorithm, we perform a ``reference'' DRB experiment (upper panel) using the parameters computed by the ESC controller (lower panel). At the first iteration of the control, we introduce an artificial displacement of the spin phase parameters $\psi_1,\,\psi_2$ from their nominal values, and see the controller correctly eliminate this offset. Additionally, the controller tracks a real underlying drift in the $\chi$ parameter over the course of the data run.}
\label{fig:exp}
\end{figure}

We validate the ESC algorithm experimentally using an IonQ trapped-ion quantum computer.
As described in Ref.~\cite{chen2023forte}, a linear chain of $^{171}$Yb$^+$ ions is trapped on a surface-electrode trap, with the qubit defined between the two $|m_F=0\rangle$ hyperfine levels of the ground state. Dynamically steerable pairs of tightly focused 355~nm laser beams drive a two-photon Raman transition to implement single-- and two-qubit gate operations. Phase, frequency, and amplitude control of the Raman laser beams is achieved via programmable RF signals sent to AOMs inserted along the beam path. This provides direct control over the Hamiltonian-level parameters that describe the unitary evolution of each quantum gate.

The two-qubit (MS) gate used in this demonstration is produced by a robust, amplitude-modulated pulse that decouples the entangled qubit states from the motional states of the ion chain at the end of the gate operation~\cite{Blmel2021}.
The amplitude-modulated envelope is designed offline, but it is parameterized by an overall scale factor (which sets $\chi$), a detuning from the qubit transition, and phase offsets of the red- and blue-sideband (RSB, BSB) tones used to generate the MS Hamiltonian. 
Specifically, our ESC controller modifies a pulse envelope scale factor on each qubit via $g_1,\, g_2$, and the RSB/BSB phase offsets via $\psi_1,\,\psi_2$.

In Fig.~\ref{fig:exp}, we show experimental results over eight iterations of the ESC algorithm. 
Before the first iteration, we displace the two spin phase controls and observe the controller drive these back to their optimal values. 
There is an additional drift in the $\chi$ parameter over the course of the data run, which the ESC algorithm correctly tracks by adjusting $g_1 g_2$. 
As an out-of-loop probe of the controller performance, we perform a ``reference'' DRB experiment to probe the gate error as the three control parameters are updated by ESC. 
For this demonstration, the ESC algorithm uses a DRB design with number of ciruit per RB depth = 4, number of shots per circuit =100, and $N_t=25$, sampled at RB depths of $[1, 50]$ and a two-qubit gate fraction of 0.75. This choice balances certain classical overheads in the experiment with noise in evaluating the RB objective function. The total runtime of each ESC iteration is $\approx9.5$~minutes.

\section{Conclusions}\label{sec:conclusion}
The extremum-seeking control strategy demonstrated in this work provides another tool for calibrating quantum systems with unknown drift. It does not rely on detailed models of the physical hardware, instead operating with a generic cost function (DRB) that is applicable to any quantum computing hardware.
The ESC procedure enables several control parameters to be extracted simultaneously under drift and noise, making this an efficient strategy for optimizing gate performance in a model-free way. While the absolute execution time is longer than what can be achieved using detailed microscopic models, this strategy may still find utility to polish gate errors after model-based calibration routines have been applied. Future work might involve optimizing the ESC algorithm over different objective functions, or comparing ESC to other strategies such as Nelder-Mead, Bayesian optimization, or simultaneous perturbation stochastic approximation.

\section{Acknowledgements}
We gratefully acknowledge support for this project from the Advancing Quantum-Enabled Technologies (AQET) traineeship program supported by NSF Award 202154 and QuantumX at the University of Washington.

\bibliographystyle{IEEEtran}
\bibliography{IEEEabrv,ref}

\begin{thebibliography}{10}
\providecommand{\url}[1]{#1}
\csname url@samestyle\endcsname
\providecommand{\newblock}{\relax}
\providecommand{\bibinfo}[2]{#2}
\providecommand{\BIBentrySTDinterwordspacing}{\spaceskip=0pt\relax}
\providecommand{\BIBentryALTinterwordstretchfactor}{4}
\providecommand{\BIBentryALTinterwordspacing}{\spaceskip=\fontdimen2\font plus
\BIBentryALTinterwordstretchfactor\fontdimen3\font minus
  \fontdimen4\font\relax}
\providecommand{\BIBforeignlanguage}[2]{{%
\expandafter\ifx\csname l@#1\endcsname\relax
\typeout{** WARNING: IEEEtran.bst: No hyphenation pattern has been}%
\typeout{** loaded for the language `#1'. Using the pattern for}%
\typeout{** the default language instead.}%
\else
\language=\csname l@#1\endcsname
\fi
#2}}
\providecommand{\BIBdecl}{\relax}
\BIBdecl

\bibitem{khaneja2005optimal}
N.~Khaneja, T.~Reiss, C.~Kehlet, T.~Schulte-Herbr{\"u}ggen, and S.~J. Glaser,
  ``Optimal control of coupled spin dynamics: design of nmr pulse sequences by
  gradient ascent algorithms,'' \emph{Journal of magnetic resonance}, vol. 172,
  no.~2, pp. 296--305, 2005.

\bibitem{motzoi2009simple}
F.~Motzoi, J.~M. Gambetta, P.~Rebentrost, and F.~K. Wilhelm, ``Simple pulses
  for elimination of leakage in weakly nonlinear qubits,'' \emph{Physical
  review letters}, vol. 103, no.~11, p. 110501, 2009.

\bibitem{blumel2021power}
R.~Bl{\"u}mel, N.~Grzesiak, N.~Pisenti, K.~Wright, and Y.~Nam, ``Power-optimal,
  stabilized entangling gate between trapped-ion qubits,'' \emph{npj Quantum
  Information}, vol.~7, no.~1, p. 147, 2021.

\bibitem{frees2019adiabatic}
A.~Frees, S.~Mehl, J.~K. Gamble, M.~Friesen, and S.~Coppersmith, ``Adiabatic
  two-qubit gates in capacitively coupled quantum dot hybrid qubits,''
  \emph{npj Quantum Information}, vol.~5, no.~1, p.~73, 2019.

\bibitem{egger2014adaptive}
D.~J. Egger and F.~K. Wilhelm, ``Adaptive hybrid optimal quantum control for
  imprecisely characterized systems,'' \emph{Physical review letters}, vol.
  112, no.~24, p. 240503, 2014.

\bibitem{kelly2014optimal}
J.~Kelly, R.~Barends, B.~Campbell, Y.~Chen, Z.~Chen, B.~Chiaro, A.~Dunsworth,
  A.~G. Fowler, I.-C. Hoi, E.~Jeffrey \emph{et~al.}, ``Optimal quantum control
  using randomized benchmarking,'' \emph{Physical review letters}, vol. 112,
  no.~24, p. 240504, 2014.

\bibitem{ferrie2015robust}
C.~Ferrie and O.~Moussa, ``Robust and efficient in situ quantum control,''
  \emph{Physical Review A}, vol.~91, no.~5, p. 052306, 2015.

\bibitem{ariyur2003real}
K.~B. Ariyur and M.~Krstic, \emph{Real-time optimization by extremum-seeking
  control}.\hskip 1em plus 0.5em minus 0.4em\relax John Wiley \& Sons, 2003.

\bibitem{proctor2019direct}
T.~J. Proctor, A.~Carignan-Dugas, K.~Rudinger, E.~Nielsen, R.~Blume-Kohout, and
  K.~Young, ``Direct randomized benchmarking for multiqubit devices,''
  \emph{Physical review letters}, vol. 123, no.~3, p. 030503, 2019.

\bibitem{Wineland1998}
D.~Wineland, C.~Monroe, W.~Itano, D.~Leibfried, B.~King, and D.~Meekhof,
  ``Experimental issues in coherent quantum-state manipulation of trapped
  atomic ions,'' \emph{Journal of Research of the National Institute of
  Standards and Technology}, vol. 103, no.~3, p. 259, May 1998.

\bibitem{Srensen1999}
A.~S{\o}rensen and K.~M{\o}lmer, ``Quantum computation with ions in thermal
  motion,'' \emph{Physical Review Letters}, vol.~82, no.~9, pp. 1971--1974,
  Mar. 1999.

\bibitem{Shantanu_thesis}
S.~Debnath, ``A programmable five qubit quantum computer using trapped atomic
  ions,'' Ph.D. dissertation, University of Maryland, 2016.

\bibitem{chen2023forte}
J.-S. Chen, E.~Nielsen, M.~Ebert, V.~Inlek, K.~Wright, V.~Chaplin, A.~Maksymov,
  E.~P{\'a}ez, A.~Poudel, P.~Maunz, and J.~Gamble, ``Benchmarking a trapped-ion
  quantum computer with 29 algorithmic qubits,'' \emph{arXiv
  preprint:2308.05071}, 2023.

\bibitem{Blmel2021}
R.~Bl\"{u}mel, N.~Grzesiak, N.~Pisenti, K.~Wright, and Y.~Nam, ``Power-optimal,
  stabilized entangling gate between trapped-ion qubits,'' \emph{npj Quantum
  Information}, vol.~7, no.~1, Oct. 2021.

\end{thebibliography}

\end{document}